\documentclass[]{raa}
\usepackage{graphicx,times}
\usepackage{natbib}
\usepackage{amssymb,amsmath}
\bibpunct{(}{)}{;}{a}{}{,}

\begin{document}

   \title{CSO CO (2-1) AND SPITZER IRAC OBSERVATIONS OF A BIPOLAR OUTFLOW IN HIGH-MASS STAR-FORMING REGION IRAS 22506+5944}


   \author{Zeqiang Xie\inst{1,2}, Keping Qiu\inst{1,2}}

   \institute{School of Astronomy and Space Science, Nanjing University, 163 Xianlin Avenue, Nanjing 210023, China; {\it kpqiu@nju.edu.cn}
   \and Key Laboratory of Modern Astronomy and Astrophysics (Nanjing University), Ministry of Education, Nanjing 210023, China
}

\abstract{
We present Caltech Submillimeter Observatory CO (2-1) and {\it Spitzer} IRAC observations toward IRAS 22506+5944, which is a 10$^4$ $L_{\odot}$ massive star-forming region. The CO (2-1) maps show an east-west bipolar molecular outflow originating from the 3 mm dust continuum peak. The {\it Spitzer} IRAC color-composite image reveals a pair of bow-shaped tips which are prominent in excess 4.5 ${\mu}$m emission and are located to the leading fronts of the bipolar outflow, providing compelling evidence for the existence of bow-shocks as the driving agents of the molecular outflow. By comparing our CO (2-1) observations with previously published CO (1-0) data, we find that the CO (2-1)/(1-0) line ratio increases from low ($\sim$5 km s$^{-1}$) to moderate ($\sim$8--12 km s$^{-1}$) velocities, and then decreases at higher velocities. This is qualitatively consistent with the scenario that the molecular outflow is driven by multiple bow-shocks. We also revisit the position-velocity diagram of the CO (1-0) data, and find two spur structures along the outflow axis, which are further evidence for the presence of multiple jet bow-shocks. Finally, power-law fittings to the mass spectrum of the outflow gives power law indexes more consistent with the jet bow-shock model than the wide-angle wind model. 
}

   \authorrunning{Z. Xie \& K. Qiu}            
   \titlerunning{CSO and Spitzer observations of IRAS 22506+5944}  
   \maketitle

%
\section{Introduction}           
\label{sect:intro}
After the first detection of a remarkably extended and collimated outflow in L1551 more than 35 years ago \citep{Snell+etal+1980}, an ever accumulating number of studies have shown that outflows are ubiquitous in both low-mass and high-mass star-forming regions (e.g., \citealt{Bachiller+Tafalla+1999}; \citealt{Zhang+etal+2001}; \citealt{Beuther+etal+2002}; \citealt{Wu+etal+2004}; \citealt{Zhang+etal+2005, Zhang+etal+2007}; \citealt{Qiu+etal+2007}; \citealt{Qin+etal+2008}; \citealt{Qiu+Zhang+2009}; \citealt{Hirano+etal+2010}; \citealt{Cyganowski+etal+2011}; \citealt{Qiu+etal+2011}; \citealt{Arce+etal+2013}; \citealt{Plunkett+etal+2013}; \citealt{Dunham+etal+2014}; \citealt{Frank+etal+2014}; \citealt{Liu+etal+2016}). The opening angle of outflows may gradually increase as the central source evolves, thereby clearing material from their parent dense cores \citep{Arce+Sargent+2006} and finally terminating the gas infall (\citealt{Velusamy+Langer+1998}). With these processes outflows play an important role in determining the star formation efficiency of a cloud (\citealt{Matzner+McKee+2000}; \citealt{Nakamura+Li+2007}; \citealt{Machida+Hosokawa+2013})  and the final mass of stars \citep{Myers+2008}. Outflows produce high velocity and energetic shocks, and thus can also significantly alter the density structure and chemistry of their parent cores, clumps, and even clouds (\citealt{Frank+etal+2014}; \citealt{Plunkett+etal+2015}). 

Outflows can be observed at various wavelengths, such as ro-vibrational lines of H$_2$ in the near-infrared (NIR) and rotational transitions of CO and some other molecules (e.g., SiO) at millimeter and submillimeter wavelengths. The NIR H$_2$ emission is attributed to shocks (including both leading and internal shocks) and often shows a bow-shaped structure along the outflow axis (e.g. \citealt{Qiu+etal+2008}; \citealt{Cyganowski+etal+2009}; \citealt{Froebrich+etal+2011}). CO outflows, especially those observed in low-J lines, are thought to consist of ambient gas being entrained or swept-up by underlying jets/winds. These molecular lines, e.g. CO (1-0) and (2-1), are easily excited compared to NIR H$_2$ lines, and are often used to measure the morphology and kinematics of molecular outflows. Regardless of numerous observational studies of molecular outflows, it is still unclear how molecular outflows are driven or accelerated. Most prevalent outflow models include jet-driven bow-shocks (leading and/or internal) (e.g., \citealt{Raga+Cabrit+1993}) and wide-angle wind-driven shells (e.g., \citealt{Li+Shu+1996}).

IRAS 22506+5944 (hereafter I22506) has been proposed as a precursor of ultra-compact H{\scriptsize II} (UC H{\scriptsize II}) regions by \cite{Molinari+etal+1996, Molinari+etal+1998}, which has a far-IR luminosity of $2.2\times10^4$ $L_{\odot}$ at an inferred distance of 5.7 kpc \citep{Su+etal+2004}. However, in spite of its high luminosity, no radio emission at 6 cm was detected at a 3$\sigma$ upper limit of $\sim$0.3 mJy beam$^{-1}$ in a $\sim$3\arcsec~ beam \citep{Molinari+etal+1998}. The detection of water maser activity and dense molecular gas, together with the properties mentioned above, makes I22506 a credible candidate for a high-mass protostar (\citealt{Wouterloot+etal+1993}; \citealt{Bronfman+etal+1996}; \citealt{Molinari+etal+1996}; \citealt{Migenes+etal+1999}). The detection of an outflow in I22506 has been reported in CO (2-1) surveys using the NRAO 12 m telescope (\citealt{Zhang+etal+2001}; \citealt{Wu+etal+2005}). \cite{Su+etal+2004} conducted followup observations in CO, $^{13}$CO, and C$^{18}$O (1-0) and continuum at 3 mm with the Berkeley-Illinois-Maryland Association (BIMA) array; the CO and $^{13}$CO (1-0) data were combined with the NRAO 12 m observations to recover the missing short-spacing information. The CO maps with a $\sim$11$''$ resolution reveals a moderately collimated, high velocity, massive, and bipolar outflow centered on the dust and gas condensation.

Here we present CO (2-1) mapping observations made with the Caltech Submillimeter Observatory (CSO) and 3--8 ${\mu}m$ imaging observations made with the {\it Spitzer} InfraRed Array Camera (IRAC) toward I22506. To investigate excitation conditions of the outflow gas and then shed light on the driving mechanism of the outflow, we jointly analyze the CSO CO (2-1) and BIMA+NRAO 12m CO (1-0) data. We search for IR counterparts of the outflow from the sensitive IRAC image, aimed at finding possible driving agents (jets or winds) of the molecular outflow.


\section{Observations}
\label{sect:Obs}
The CO (2-1) observations were made with the CSO on 2009 June 4. The output signal from a 230 GHz receiver was processed by a Fast Fourier Transform Spectrometer which has a total bandwidth of 500 MHz divided into 8192 channels. The weather conditions were excellent during the observations, with the atmospheric opacity at 225 GHz, $\tau_{\rm 225 GHz}$, around 0.05, and the system temperature is about 400 K. We made mapping observations with the on-the-fly (OTF) mode, and obtained a $15\times15$ grid map with a grid cell size of $15''$, corresponding to a $\sim4'\times4'$ map centered on (R.A., Decl.)$_{\rm J2000}$=(22$^h$52$^m$38$^s$.17, 60$^\circ$00\arcmin50\arcsec.30). The effective on-source integration time on each grid cell is 10 seconds. The data were processed with the GILDAS/CLASS package for baseline fitting and subtraction, and velocity smoothed into 1.27 km s$^{-1}$ channels. Unless specified, the data are presented in $T_{\rm A}^\ast$ and have an RMS sensitivity of 0.14 K. A conversion from $T_{\rm A}^\ast$ to $T_{\rm mb}$, whenever needed, could be derived with a main beam efficiency of 0.7, following http://www.submm.caltech.edu/cso/receivers/beams.html.

The {\it Spitzer} IRAC \citep{Fazio+etal+2004} observations were obtained from the {\it Spitzer} archive (Program ID: 50264). The observations were made in the High Dynamic Range mode with integration times of 0.4 and 10.4 seconds per dither, and 16 dithers in total, resulting in a total effective integration time of 166.4 seconds per pixel. The short integration (0.4 second) frames are crucial to image bright sources without saturation, and the long integration (10.4 seconds) frames could image faint structures. The mapping area covered by all four bands is roughly $5'\times5'$. The frames were processed by the {\it Spitzer} Science Center for with the standard pipeline version S18.7 to produce Post Basic Calibrated Data products.

\section{Results}
\label{sect:Results}

\subsection{CO (2-1) Emission and {\it Spitzer} IRAC Image}
\label{subsect:CO2-1andSpitzers}

\begin{figure}
   \centering
  \includegraphics[width=\textwidth, angle=0]{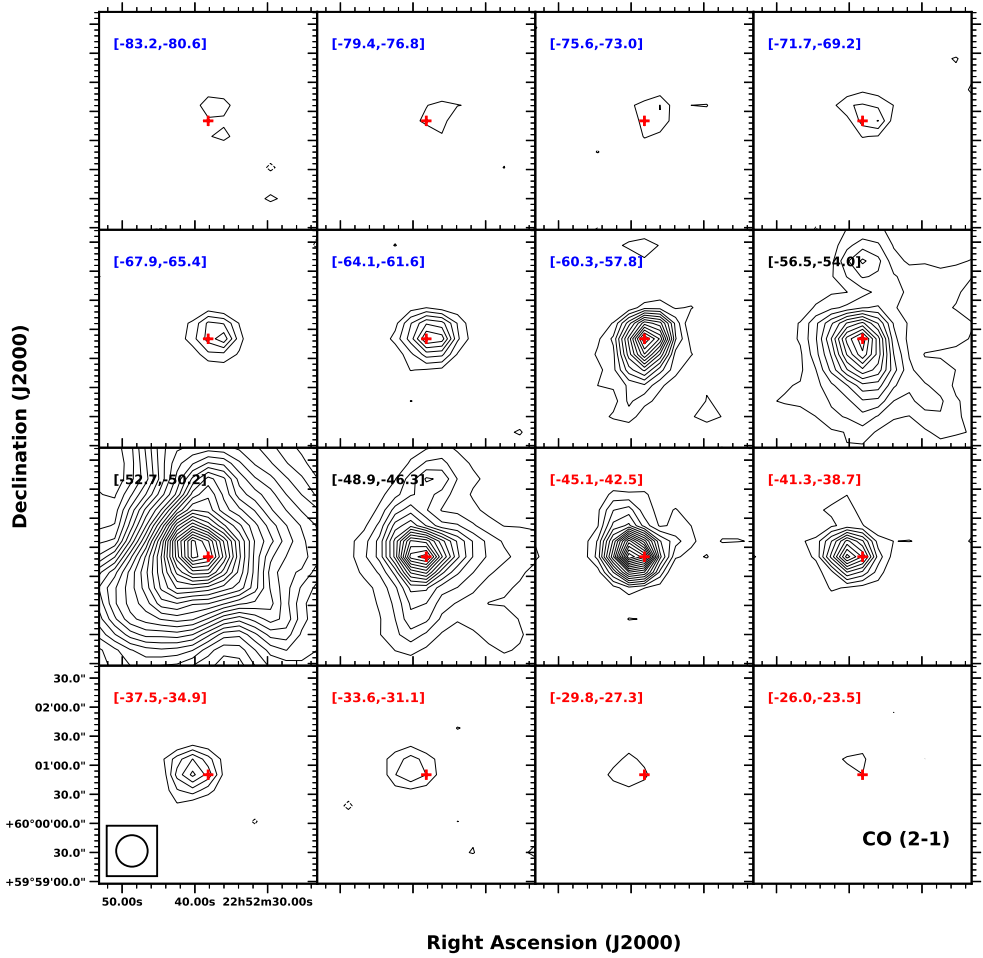}
   \caption{Contour maps of the CO (2-1) emission integrated over every three channels, with the velocity range of each panel given in the upper left corner. Solid/dashed contours represent positive/negative emissions starting at and continuing in steps of $\pm3\sigma$, where $\sigma=0.3$ K km s$^{-1}$. For the central three panels with velocities close to $V_{\rm cloud}$ (from $-56.5$ to $-46.3$ km s$^{-1}$), the CO emission is heavily affected by the ambient molecular cloud, and the lowest and stepping contour levels are set to $\pm6\sigma$. A plus symbol depicts the 3 mm continuum peak from \cite{Su+etal+2004}. The CSO beam size is shown in the lower left panel.}
   \label{fig:channelmap}
   \end{figure}

\begin{figure}
   \centering
  \includegraphics[width=\textwidth, angle=0]{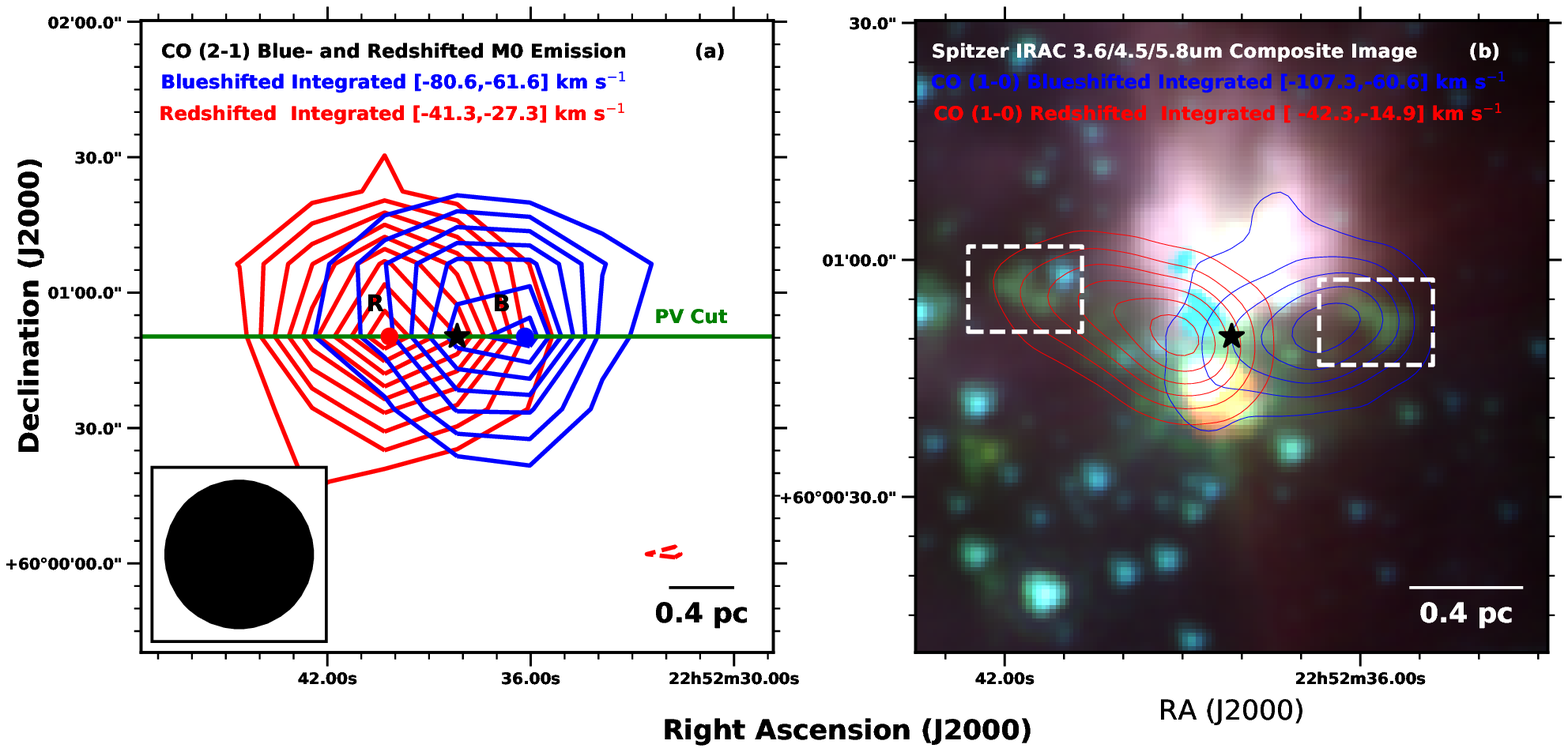}
   \caption{(a) Blue and red contours show the CO (2-1) emissions integrated from $-80.6$ to $-61.6$ km s$^{-1}$ and from $-41.3$ to $-27.3$ km s$^{-1}$, respectively, with contour levels starting at $4\sigma$ and continuing in steps of $3\sigma$, where $\sigma=0.7$ K km s$^{-1}$ for the blue lobe and 0.6 K km s$^{-1}$ for the red lobe. Two dots, namely ``B'' and ``R'', mark the peaks of the two lobes. A star symbol represents the 3 mm continuum peak. A green line going through the peaks of the outflow lobes approximately intersects the 3 mm continuum peak, and also denotes the cut used for a position-velocity (PV) diagram. A filled circle in the lower left corner shows the CSO  beam. (b) {\it Spitzer} three-color composite image with the 3.6, 4.5 and 5.8 ${\mu}$m emissions coded in blue, green, and red, respectively. Two dashed rectangles outline a pair of bow-shaped tips prominent in the 4.5 ${\mu}$m band (green). Blue and red contours show the CO (1-0) emissions integrated from $-107.3$ to $-60.6$ km s$^{-1}$ and from $-42.3$ to $-14.9$ km s$^{-1}$, respectively, with contour levels starting at 15 K km s$^{-1}$ and continuing in steps of 7.5 K km s$^{-1}$ for the blue lobe, and starting at 11 K km s$^{-1}$ and continuing in steps of 5.5 K km s$^{-1}$ for the red lobe. A filled triangle in the center marks the 3 mm continuum peak.}
   \label{fig:integrated}
   \end{figure}

Figure~\ref{fig:channelmap} presents the velocity channel maps of the CO (2-1) emission, where the cloud systemic velocity ($V_{\rm cloud}$) with respect to the local standard of rest (LSR) is about $-51.5$ km s$^{-1}$ \citep{Su+etal+2004}. For presentation purposes, we have integrated every three channels into a wide channel with a width of 3.81 km s$^{-1}$. The CO (2-1) emission arising from the outflow is detected as high velocity structures reaching $\sim-80$ km s$^{-1}$ for the blueshifted lobe and $\sim-26$ km s$^{-1}$ for the redshifted lobe. The emission near the systemic velocity (within $V_{\rm cloud}\pm5$ km s$^{-1}$) fills almost the entire field of view, and mostly traces the ambient molecular cloud. The high velocity blueshifted emission has its peak located to the west of the 3 mm continuum peak, and the redshifted emission is detected to the east; thus we are detecting an east-west, bipolar molecular outflow centered on the dust and gas condensation, all consistent with the CO (1-0) observations from \cite{Su+etal+2004}. An overall picture of the bipolar outflow is shown in Figure~\ref{fig:integrated}(a), where the CO (2-1) emission is integrated over high velocity line wings from $-80.6$ to $-61.6$ km s$^{-1}$ for the blueshifted lobe and from $-41.3$ to $-27.3$ km s$^{-1}$ for the redshifted lobe.

Figure~\ref{fig:integrated}(b) shows the {\it Spitzer} IRAC image overlaid with the high-resolution contour map of the velocity integrated CO (1-0) emission. The IRAC image is made with emissions in the 3.6, 4.5, and 5.8 $\mu$m bands coded in blue, greed, and red; such a color-composite image has been proven to be a useful diagnostic tool for shocked H$_2$ emission in outflows (e.g. \citealt{Raga+etal+2004}; \citealt{Noriega+etal+2004}; \citealt{Smith+etal+2006}; \citealt{Qiu+etal+2008}; \citealt{Cyganowski+etal+2009}). Even though contributions from other lines (e.g., Br$\gamma$, CO band head) cannot be ruled out without spectroscopic observations, in many cases, excess 4.5 $\mu$m emission arising from outflows is mostly due to shocked H$_2$ lines (\citealt{Smith+Rosen+2005}; \citealt{Smith+etal+2006}; \citealt{De+Vacca+2010}). In Figure~\ref{fig:integrated}(b), we detect a pair of bow-shaped tips prominent in the 4.5 $\mu$m band (green); the two tips pointing to the east and west, having an orientation approximately consistent with that of the CO outflow, and are located to the leading fronts of the molecular outflow. Apparently the 4.5 $\mu$m tips are tracing leading bow-shocks produced by a fast jet emanating from the central dust and gas condensation; the jet head has traveled further away to a less dense medium, leaving behind a bipolar molecular outflow seen in CO lines. 

\subsection{Line Ratio of CO (2-1)/(1-0) and Excitation Temperature of the Outflow}
\label{subsect:lineratio}

\begin{figure}
   \centering
  \includegraphics[width=\textwidth, angle=0]{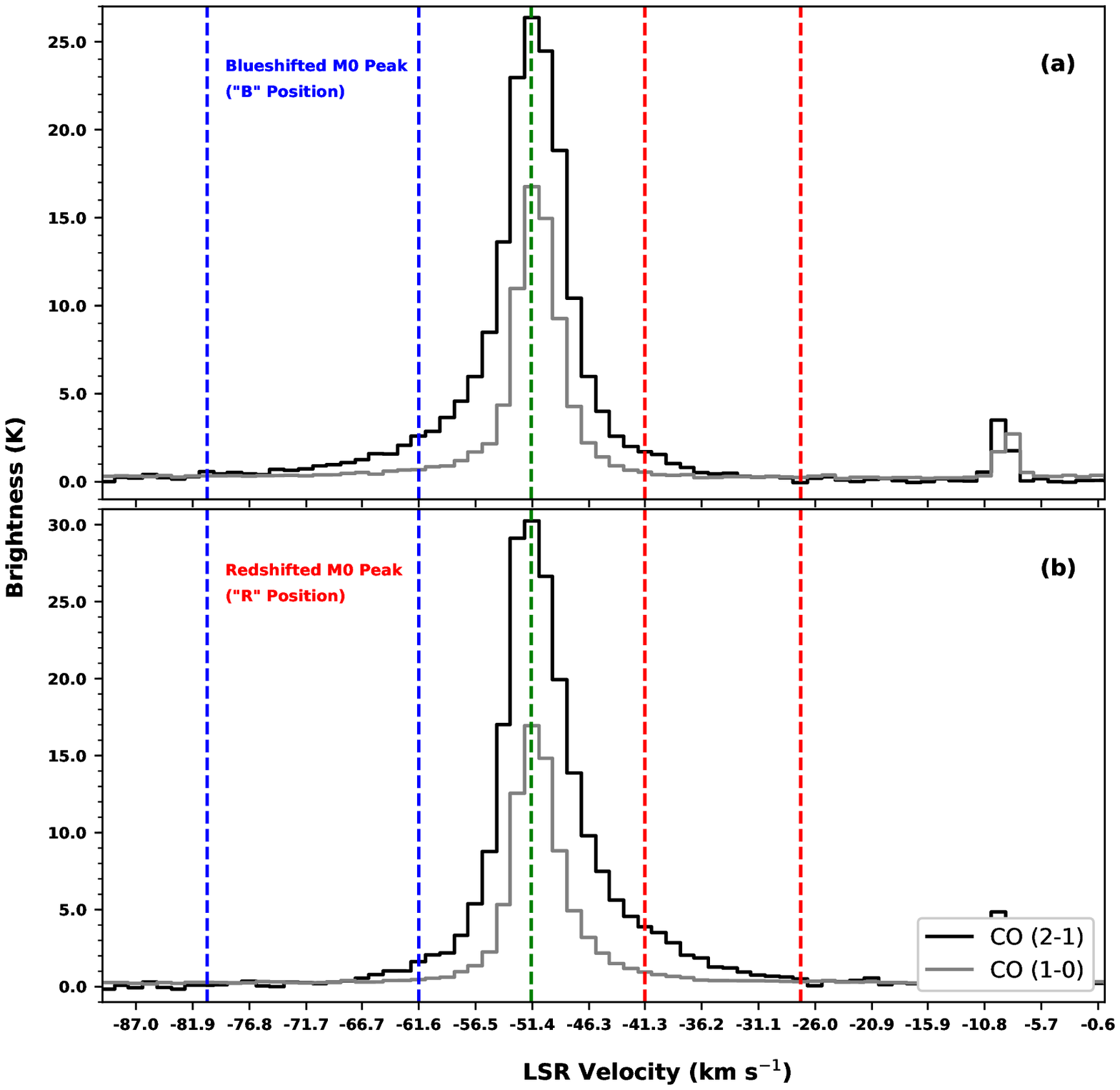}
   \caption{(a) CO (2-1) and CO (1-0) spectra extracted from the peak of the blueshifted lobe (``B'' position in Figure~\ref{fig:integrated}(a)) shown in black and gray histograms, respectively. The CO (2-1) data have been converted from $T_{\rm A}^\ast$ to $T_{\rm mb}$. The CO (1-0) data have been convolved and resampled to match the CO (2-1) spatial and spectral resolutions. A vertical dashed line in green indicates $V_{\rm cloud}$, and vertical dashed lines in blue/red mark the blueshifted/redshifted velocity intervals for the map shown in Figure~\ref{fig:integrated}(a). (b) Same as (a), but extracted from the peak of redshifted lobe (``R'' position in Figure~\ref{fig:integrated}(a)).}
   \label{fig:spectrum}
   \end{figure}

\begin{figure}
   \centering
  \includegraphics[width=14.5cm, angle=0]{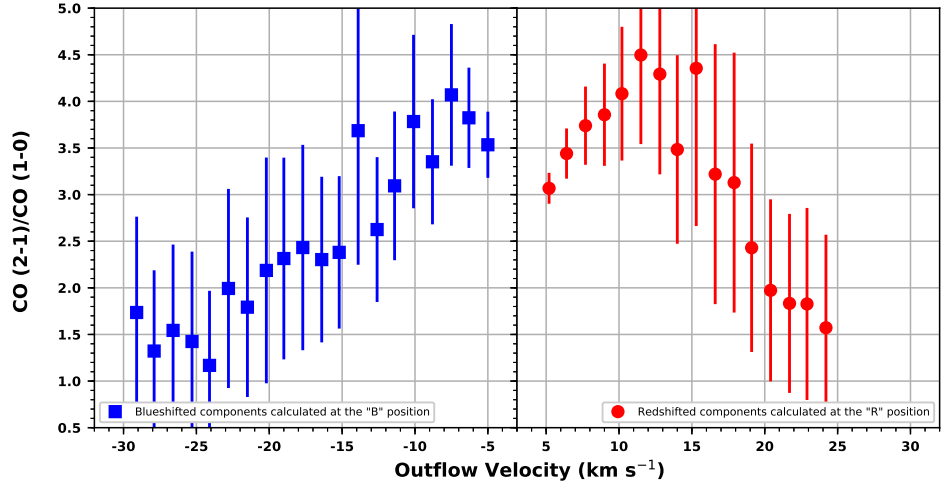}
   \caption{The line ratio of CO (2-1) to CO (1-0) as a function of outflow velocity. The blue filled squares and red filled circles denote the line ratio calculated from the peak of the CO (2-1) blue- and redshifted lobe, repectively. The error bars indicate the 1$\sigma$ error. Assuming very optically thin limit for both CO lines, the excitation temperatures of 10, 20, 30, 40 and 50 K implied by the line ratios are marked with dotted lines, as well as the error shown in gray shaded area.}
   \label{fig:ratio}
   \end{figure}
   
\begin{table}
\bc
\begin{minipage}[]{145mm}
\caption[]{Measured CO (2-1) and (1-0) brightness, and calculated CO (2-1) / (1-0) line ratios ($R_{21/10}$) and excitation temperatures ($T_{\rm ex}$) at each outflow velocity for both blueshifted and redshifted lobes \label{tab:ratio}}\end{minipage}
\setlength{\tabcolsep}{2.5pt}
\small
 \begin{tabular}{ccccccccccc}
  \hline\noalign{\smallskip}
   \multicolumn5c{Blueshifted Lobe} & \multicolumn5c{Redshifted Lobe}  \\
  ~~$v_{\rm out}$~~ & ~~CO (2-1)~~ & ~~CO (1-0)~~ & ~~~$R_{21/10}$~~~ & ~~~$T_{\rm ex}$~~~ & & ~~$v_{\rm out}$~~ & ~~CO (2-1)~~ & ~~CO (1-0)~~ & ~~~$R_{21/10}$~~~ & ~~~$T_{\rm ex}$~~~ \\
  (km~s$^{-1}$) & (K) & (K) & & (K) & & (km~s$^{-1}$) & (K) & (K) &  & (K) \\
  \hline\noalign{\smallskip}
5.0 & 5.98 & 1.69 & 3.53 $\pm$ 0.35 & 88.9 &  & 5.2 & 9.80 & 3.19 & 3.07 $\pm$ 0.17 & 41.5 \\
6.3 & 4.58 & 1.20 & 3.82 $\pm$ 0.54 & 244.1 &  & 6.4 & 7.49 & 2.18 & 3.44 $\pm$ 0.27 & 73.0 \\
7.5 & 3.65 & 0.90 & 4.07 $\pm$ 0.76 & $\geq58.1$ & & 7.7 & 5.62 & 1.50 & 3.74 $\pm$ 0.42 & 163.3 \\
8.8 & 2.86 & 0.85 & 3.35 $\pm$ 0.67 & 62.3 &  & 9.0 & 4.56 & 1.18 & 3.86 $\pm$ 0.55 & 301.6 \\
10.1 & 2.59 & 0.69 & 3.78 $\pm$ 0.93 & 197.8 &  & 10.2 & 3.89 & 0.95 & 4.08 $\pm$ 0.72 & $\geq63.6$ \\
11.4 & 2.07 & 0.67 & 3.09 $\pm$ 0.80 & 42.8 &  & 11.5 & 3.51 & 0.78 & 4.50 $\pm$ 0.96 & $\geq90.4$ \\
12.6 & 1.57 & 0.60 & 2.63 $\pm$ 0.78 & 26.1 &  & 12.8 & 2.86 & 0.67 & 4.29 $\pm$ 1.08 & $\geq50.5$ \\
13.9 & 1.60 & 0.43 & 3.69 $\pm$ 1.44 & 134.3 &  & 14.0 & 2.04 & 0.59 & 3.48 $\pm$ 1.01 & 79.6 \\
15.2 & 1.25 & 0.53 & 2.38 $\pm$ 0.82 & 21.2 &  & 15.3 & 1.87 & 0.43 & 4.36 $\pm$ 1.69 & $\geq27.0$ \\
16.4 & 1.09 & 0.47 & 2.30 $\pm$ 0.89 & 19.9 &  & 16.6 & 1.28 & 0.40 & 3.22 $\pm$ 1.39 & 50.7 \\
17.7 & 0.97 & 0.40 & 2.43 $\pm$ 1.10 & 22.1 &  & 17.9 & 1.21 & 0.39 & 3.13 $\pm$ 1.39 & 44.8 \\
19.0 & 0.90 & 0.39 & 2.31 $\pm$ 1.08 & 20.1 &  & 19.1 & 0.95 & 0.39 & 2.43 $\pm$ 1.12 & 22.1 \\
20.2 & 0.73 & 0.33 & 2.19 $\pm$ 1.21 & 18.2 &  & 20.4 & 0.76 & 0.38 & 1.97 $\pm$ 0.98 & 15.6 \\
21.5 & 0.65 & 0.36 & 1.79 $\pm$ 0.96 & 13.7 &  & 21.7 & 0.68 & 0.37 & 1.83 $\pm$ 0.96 & 14.1 \\
22.8 & 0.70 & 0.35 & 1.99 $\pm$ 1.07 & 15.8 &  & 22.9 & 0.63 & 0.34 & 1.83 $\pm$ 1.03 & 14.0 \\
24.1 & 0.40 & 0.34 & 1.17 $\pm$ 0.80 & 8.9 &  & 24.2 & 0.51 & 0.32 & 1.57 $\pm$ 1.00 & 11.8 \\
25.3 & 0.45 & 0.31 & 1.42 $\pm$ 0.96 & 10.7 & & & & & \\
26.6 & 0.53 & 0.35 & 1.54 $\pm$ 0.92 & 11.6 & & & & & \\
27.9 & 0.44 & 0.34 & 1.32 $\pm$ 0.87 & 9.9 & & & & & \\
29.1 & 0.58 & 0.33 & 1.74 $\pm$ 1.03 & 13.2 & & & & & \\
  \noalign{\smallskip}\hline
\end{tabular}
\ec
\tablecomments{\textwidth}{
The brightness of the two lines is measured at the peaks of the two lobes of the CO (2-1) outflow, which are marked as "B" and "R" positions in Figure~\ref{fig:integrated}(a).}
\end{table}    

In order to calculate the line ratio of CO (2-1) to (1-0) and to investigate excitation conditions of the outflow gas as a function of the velocity, we reconstruct the combined BIMA and NRAO 12 m CO (1-0) data from \citet{Su+etal+2004}: the CO (1-0) map was convolved from a beam size of $11.\!''4\times10.\!''7$ to $32.\!''5\times32.\!''5$, and the velocity axis was resampled from a resolution of 1.02 km s$^{-1}$ to 1.27 km s$^{-1}$. Figure~\ref{fig:spectrum} shows the CO (2-1) and (1-0) spectra extracted from the peaks of the blueshifted and redshifted lobes (marked as ``B'' and ``R" in Figure~\ref{fig:integrated}(a)). The spectral line profiles and in particular the high velocity line wings of the two transitions are consistent with each other. We then derive the line ratio of CO (2-1) to (1-0), $R_{21/10}$, at each velocity channel, which is listed in Table~\ref{tab:ratio} along with the measured intensities of the two lines.

Assuming local thermodynamical equilibrium (LTE) and optically thin emissions for the CO (2-1) and (1-0) lines, the excitation temperature, $T_{\rm ex}$, could be derived from the line ratio following $R_{21/10} = 4e^{-11/T_{\rm ex}}$.  \citet{Su+etal+2004} compared the CO (1-0) to $^{13}$CO (1-0) emissions to estimate the opacity of the CO (1-0) line, and found that the line is optically thick for the velocity range of $V_{\rm cloud}\pm5$ km s$^{-1}$; at higher velocities, the line was assumed to be optically thin since the $^{13}$CO (1-0) emission was not detected. We therefore calculate $T_{\rm ex}$ from $R_{21/10}$ for $v_{\rm out}\gtrsim5$ km s$^{-1}$, where $v_{\rm out}$ is the absolute difference between the LSR velocity of the outflow gas and $V_{\rm cloud}$. There are several velocity channels having line ratios greater than 4 (see Table~\ref{tab:ratio}), making a direct calculation of $T_{\rm ex}$ impossible under LTE and optically thin assumptions; we adopted a lower limit of the line ratio considering uncertainties for those channels. The derived $T_{\rm ex}$ at each velocity channel varies from $\sim10$ K to $>100$ K (see Table~\ref{tab:ratio}). The variation of excitation conditions of the outflow gas is of great interests to an investigation of the outflow driving mechanism. However, we refrain from checking the $T_{\rm ex}$ versus $v_{\rm out}$ relation for two reasons. First, as indicated above, for some channels with $R_{21/10}>4$, we could only derive a lower limit of $T_{\rm ex}$. Second, the accuracy of the $T_{\rm ex}$ estimation depends on the optical depth, and an opacity of 0.1 could lead to an underestimate by up to 20\% for $R_{21/10}<2.5$ and by up to 40\% for $R_{21/10}>2.8$ \citep{Arce+Goodman+2002}. Nevertheless, as argued by \citet{Arce+Goodman+2002}, it is true that a higher line ratio implies a higher excitation temperature even with a large opacity. We therefore show the $R_{21/10}$ versus $v_{\rm out}$ relation in Figure~\ref{fig:ratio}, and discuss the implications in Section~\ref{mechanism}.

\section{Discussion}
\label{sect:discussion}
   
\subsection{Mass and Energetics of the Outflow}
\label{subsec:outflowpropetry}   

By assuming LTE and optically thin emission for the CO (2-1) line wings, the outflow mass is calculated following $$M_{\rm out}(v_{\rm out})=5.3\times10^{-8}\,e^{16.6/T_{\rm ex}}\,(T_{\rm ex}+0.92)\,d_{\rm kpc}^2\,{\delta}s\,{\Sigma}T_{\rm mb}\,{\Delta}v,$$ where $M_{\rm out}(v_{\rm out})$ is the gas mass in a channel of $v_{\rm out}$, $d_{\rm kpc}$ is the source distance in kpc, ${\delta}s$ is the pixel size in arc~second$^2$, ${\Sigma}T_{\rm mb}$ is the main beam temperature summed over pixels with signal-to-noise ratios greater than 3, and ${\Delta}v$ is the channel width in km s$^{-1}$. For the excitation temperature, again because for some channels we only obtain a lower limit, we calculate an average $T_{\rm ex}$ for each lobe, which is 35 K for the blueshifted lobe and 66 K for the redshifted lobe. The derived outflow mass amounts to 22.7 $M_{\odot}$, which is well comparable with that of \citet{Su+etal+2004} (20 $M_{\odot}$), indicating that the gas at very high velocities ($\gtrsim25$ km s$^{-1}$), of which the emission is detected in CO (1-0) by \citet{Su+etal+2004} but not detected here in CO (2-1), contributes little to the total mass of the outflow. We then calculate the outflow momentum, $P_{\rm out}= \sum{M_{\rm out}(v_{\rm out})v_{\rm out}}$, and the kinetic energy, $E_{\rm out}=0.5\sum{M_{\rm out}(v_{\rm out})v_{\rm out}^{2}}$. The outflow radius $R$ is measured as the average distance from the peaks of the two lobes to the central source. The characteristic velocity is defined as $V=P_{\rm out}/M_{\rm out}$. Thus we can estimate the dynamic timescale of the outflow $t_{\rm dyn}=R/V$. Consequently, dynamical parameters of the outflow, such as the mass outflow rate, $\dot{M}_{\rm out}=M_{\rm out}/t_{\rm dyn}$, mechanical luminosity, $L_{\rm out}=E_{\rm out}/t_{dyn}$, and the driving force, $F_{\rm out}=P_{\rm out}/t_{dyn}$, are then computed. All the calculated outflow parameters are listed in Table~\ref{tab:outflow}. Note that we do not correct for an unknown inclination angle for the calculations, while \citet{Su+etal+2004} made a correction by adopting an inclination angle of $45^{\circ}$. If the same correction is applied to our data, parameters $R$, $V$, $P_{\rm out}$, $F_{\rm out}$ will be increased by a factor of 1.4, and parameters $E_{\rm out}$, $L_{\rm out}$ will be increased by a factor of 2, while other parameters ($M_{\rm out}$, $t_{\rm dyn}$, $\dot{M}_{\rm out}$) would remain the same. Nevertheless, all the parameters derived from our CO (2-1) observations are comparable with those derived from CO (1-0) by \citet{Su+etal+2004}, confirming that we are observing a massive and energetic outflow originating from a high-mass protostar. If we assume that the outflow is driven by an underlying jet or wind with a velocity of order 500 km s$^{-1}$ and adopting a ratio of $1/3$ for the jet or wind mass loss rate to mass accretion rate, we obtain a mass accretion rate of order $10^{-4}M_{\odot}$ yr$^{-1}$, which is sufficiently high to overcome radiation pressure from the central high-mass protostar or young star (see \citealt{Qiu+Zhang+2009} and references therein).

\begin{table}
\bc
\begin{minipage}[]{145mm}
\caption[]{Computed Outflow Prameters \label{tab:outflow}}\end{minipage}
\setlength{\tabcolsep}{2.5pt}
\small
 \begin{tabular}{lcc}
  \hline\noalign{\smallskip}
  Parameters & ~~~CO (2-1)~~~  & ~~~CO (1-0)~~~ \\
                     & ~~~this work~~~ & ~~~\citet{Su+etal+2004}~~~ \\
  \hline\noalign{\smallskip}
  Outflow Velocity   &    &  \\
  ~~~Blueshifted (km s$^{-1}$) & $5.0 \leq v_{\rm out} \leq 29.1$ & $4.0 \leq v_{\rm out} \leq 55.8$ \\
  ~~~Redshifted (km s$^{-1}$) & $5.2 \leq v_{\rm out} \leq 22.9$ & $4.1 \leq v_{\rm out} \leq 36.6$ \\
  Outflow Mass   &   &    \\
  ~~~Blueshifted, $M_{\rm blue}$ ($M_{\odot}$)   &  8.8  &  7.0 \\
  ~~~Redshifted, $M_{\rm red}$ ($M_{\odot}$)  &  13.9  &  13.0 \\
  ~~~Total, $M_{\rm out} = M_{\rm blue} + M_{\rm red} $ ($M_{\odot}$)  &  22.7  &  20.0 \\
  Outflow Momentum $P_{\rm out}$ ($M_{\odot}$ km s$^{-1}$)  &  189.0  &  180.0 \\
  Outflow Energy $E_{\rm out}$ (10$^{46}$ erg) &  1.9 & 2.5 \\
  Outflow Radius $R$ (pc)  &  0.3   &   0.2 \\
  Characteristic Velocity $V = P_{\rm out}/M_{\rm out}$ (km s$^{-1}$) & 8.3 & 9.4 \\
  Dynamical Timescale $t_{\rm dyn} = R/V $ (10$^4$ yr) & 3.8 & 1.9 \\
  Outflow Mass-Loss Rate ${\dot{M}}_{\rm out} = M_{\rm out}/t_{\rm dyn}$ (10$^{-4}$ $M_{\odot}$ yr$^{-1}$) & 6.0 & 11.0 \\
  Mechanical Luminosity $L_{\rm out} = E_{\rm out}/t_{\rm dyn}$ ($L_{\odot}$) & 4.3 & 11.0 \\
  Driving Force $F_{\rm out} = P_{\rm out}/t_{\rm dyn}$ (10$^{-3}$ $M_{\odot}$ km s$^{-1}$ yr$^{-1}$) & 5.0 & 9.5 \\
  \noalign{\smallskip}\hline
\end{tabular}
\ec
\end{table}
   
\subsection{Mass-Velocity and Position-Velocity Diagrams of the Outflow}

\begin{figure}
   \centering
  \includegraphics[width=\textwidth, angle=0]{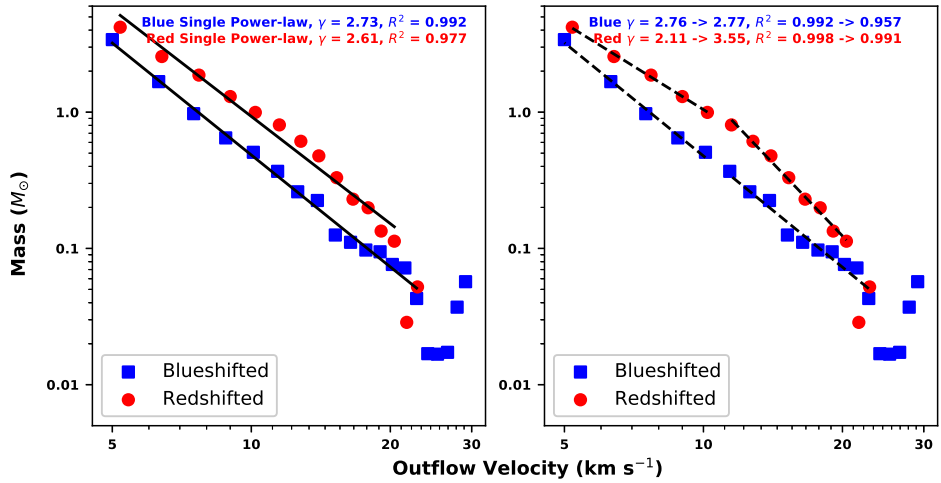}
   \caption{Mass-velocity diagram of the the CO (2-1) outflow with a single power-law fitting (a) and a broken power-law fitting (b). Blue filled squares denote the measurements from the blueshifted lobe, and red filled circles the redshifted lobe. The fitted power-law index ($\gamma$) and the corresponding Goodness of Fit ($R^2$) are given in the upper right of each panel. Data points at the highest velocities ($\gtrsim22$ km s$^{-1}$) are not included in the fittings.}
   \label{fig:m-v}
   \end{figure}

The mass-velocity relation of molecular outflows often exhibits a power law behavior, $M_{\rm out}(v_{\rm out}) \propto v_{\rm out}^{-\gamma}$, which can be a diagnostic tool for the interaction between an underlying jet or wind with the ambient gas even though its physical origin is not well established (\citealt{Chandle+etal+1996}; \citealt{Lada+Fich+1996}; \citealt{Richer+etal+2000}; \citealt{Ridge+Moore+2001}; \citealt{Su+etal+2004}; \citealt{Arce+etal+2007}; \citealt{Qiu+etal+2007}; \citealt{Qiu+Zhang+2009}; \citealt{Qiu+etal+2011}). Many observations find $\gamma$ changes at a velocity between 6 and 12 km s$^{-1}$ with a steeper power law index at higher velocity \citep{Arce+etal+2007}. In recent MHD simulations, the break velocity is found to fall in the range of 4 to 20 km s$^{-1}$, and a lower break velocity can be due to the weakness or youth of the outflow, or a large inclination angle \citep{Li+etal+2017}. Figure~\ref{fig:m-v} shows the mass-velocity (M-V) diagrams of the CO (2-1) outflow, where a broken power fitting to the redshifted lobe, with $\gamma$ steepening from 2.11 to 3.55 at about 12 km s$^{-1}$, appears to improve over a single power law fitting with $\gamma=2.61$. However, the blueshifted lobe does not show any clear trend for a broken power law; instead, it can be readily fitted with a single power law with $\gamma=2.73$, and if we try a broken power law fitting, the slopes still remain the same for $v_{\rm out}<12$ km s$^{-1}$ and $v_{\rm out}>12$ km s$^{-1}$. A single power law has also been reported for some other outflows (\citealt{Plunkett+etal+2015}; \citealt{Zhang+etal+2016}). The mass spectrum obtained from the CO (1-0) data by \citet{Su+etal+2004} shows a broken power law for both lobes, with $\gamma$ steepens at about 10 km s$^{-1}$ from 1.0 to 2.5 for the blueshifted lobe and from 2.1 to 3.2 for the redshifted lobe. Thus the mass-velocity relations derived from the CO (2-1) and (1-0) observations are consistent with each other for the redshifted lobe, but vary for the blueshifted lobe; the differences in spatial resolutions and velocity ranges may account for the variations. \cite{Lee+etal+2001} performed two-dimensional hydrodynamic simulations of jet-driven and wind-driven models, and found that the jet bow-shock model yields $\gamma$ ranging from 1.5 to 3.5, while the wide-angle wind model has $\gamma$ in a narrow range of 1.3 to 1.8. The mass spectrum of the I22506 outflow has $\gamma$ varying from 1.0 to 3.6 taking into account both the CO (2-1) and (1-0) observations, and appear to be more consistent with the jet bow-shock model.

\begin{figure}
   \centering
  \includegraphics[width=\textwidth, angle=0]{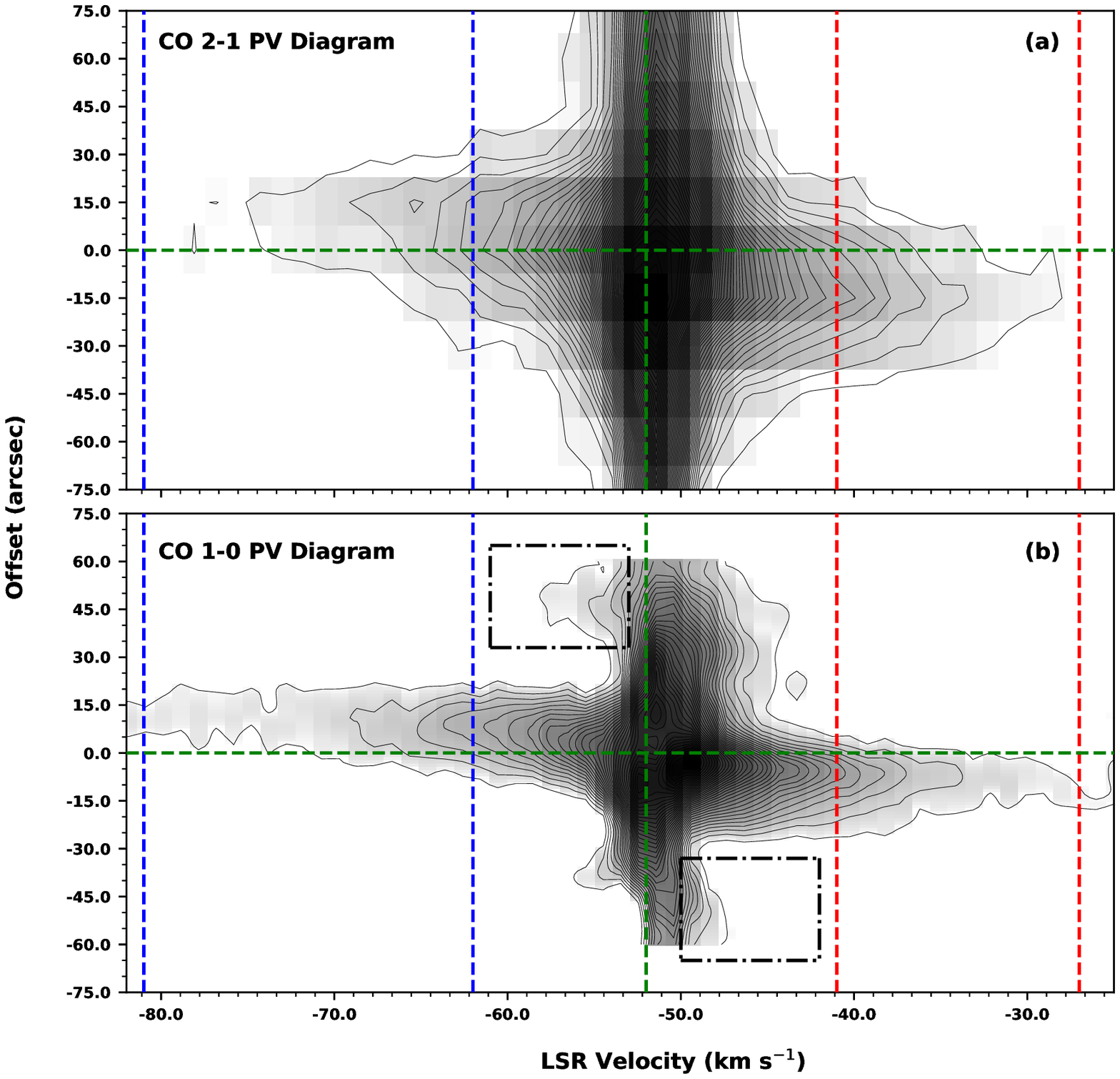}
   \caption{(a) Position-velocity (P-V) diagram of the CO (2-1) outflow, constructed along the cut shown in Figure~\ref{fig:integrated}(a). The data have been converted to the $T_{\rm mb}$ scale  for a comparison with the CO (1-0) data. The lowest and spacing contour levels are 0.6 K ($3\sigma$), and the gray scale stretches in a square-root algorithm from 0.6 K to the peak emission of 31.4 K.  The offsets are measured with respect to the 3 mm continuum peak. (b) P-V diagram in CO (1-0), constructed from the same cut as that for CO (2-1). Note the data here retain the original spatial and spectral resolutions as in \citet{Su+etal+2004}, except that the data have been converted to the brightness temperature scale. The lowest and spacing contour levels are 0.72 K, and the gray scale stretches in a square-root algorithm from 0.72 K to the peak emission of 27.5 K. Two dash-dotted rectangles outline a pair of ``spur'' structures approximately 50\arcsec away from the central source. In both panels, the vertical dashed lines are the same as those in Figure~\ref{fig:spectrum}.}
   \label{fig:p-v}
   \end{figure}

Figure~\ref{fig:p-v} shows CO (2-1) and (1-0) Position-Velocity (P-V) diagrams of the outflow, both constructed along a cut going through the peaks of the CO (2-1) blueshifted and redshifted lobes (see Figure~\ref{fig:integrated}(a)). Both diagrams are dominated by a bipolar high velocity structure. \citet{Su+etal+2004} made a P-V cut slightly different from Figure~\ref{fig:p-v}(b), revealing a high velocity structure similar to Figure~\ref{fig:p-v}(b), and furthermore, they found that the redshifted lobe shows a Hubble law, that is, the terminal velocity nearly linearly increases with the outflow velocity. A Hubble flow is consist with the jet bow-shock model (\citealt{Lee+etal+2000, Lee+etal+2001}; \citealt{Arce+etal+2007}). A Hubble law pattern is not that clear neither in Figure~\ref{fig:p-v}(a) mostly due to the lower resolution, nor in Figure~\ref{fig:p-v}(b) because of the different orientations of the cuts. However, Figure~\ref{fig:p-v}(b) shows additional features: at a distance about $50''$ from the central source, we identify a new spur-like structure in each side of the central source. According to analytical and numerical works (\citealt{Lee+etal+2000, Lee+etal+2001, Lee+etal+2002}; \citealt{Arce+Goodman+2001}), an outflow driven by an episodic jet producing multiple bow-shocks will show multiple spur structures (or Hubble wedges) in the P-V diagram. The outer spurs along with the inner very high velocity structures thus may probe multiple jet bow-shocks and furthermore an episodic nature of the mass ejection process in the central high-mass protostar. 

\subsection{Driving Mechanism of the Outflow}
\label{mechanism}

With the sensitive {\it Spitzer} IRAC image, we discover a pair of bow-shaped tips prominent in the 4.5 $\mu$m band, and the emission is presumably dominated by shocked H$_2$ lines. The two tips are located to the leading fronts of the bipolar molecular outflow, providing strong evidence for the jet bow-shock driving mechanism for the molecular outflow. The M-V and P-V relations of the molecular outflow are both consistent with the jet bow-shock model. In particular, a revisit to the high resolution CO (1-0) P-V diagram yields multiple spurs which are indicative of multiple bow-shocks as the driving agents of the molecular outflow. Comparing the distances to the central source of the bow-shaped tips seen in the IRAC image and the outer spurs in the P-V diagram, the two phenomena are likely associated with each other. In this case, the IRAC tips trace a leading bow-shock created by ejecta which has traveled to a longer distance, whereas the highest velocity gas (corresponding to the inner spur in the P-V diagram) is driven by an internal bow-shock induced by new ejecta which have just emerged from the central dust core and could not be detected in the IRAC image due to a higher extinction close to the dust core.

The excitation temperature of the outflow gas can also help to discriminate between the jet bow-shock and wide-angle wind models: in the jet bow-shock model, the gas temperature would monotonically increase with the velocity for a steady jet which produces a single leading bow-shock, but the temperature could reach the peak at an intermediate velocity for an episodic jet with multiple bow-shocks, where the leading bow-shock heats the gas to the highest temperature and the internal shock accelerates the gas to the highest velocities; in contrast, the gas temperature remains approximately constant in the wide-angle wind model (see Fig. 1 in \citealt{Arce+Goodman+2002} and Fig. 2 in \citealt{Arce+etal+2007}). With a joint analysis of the CO (2-1) and (1-0) data, we derive the line ratio as a function of the outflow velocity (Section \ref{subsect:lineratio}). In Figure \ref{fig:ratio}, the line ratio of the redshifted lobe increases from $\sim5$ km s$^{-1}$ to 12 km s$^{-1}$, and then decreases with the velocity. A temperature increase in the blueshifted lobe is not that remarkable, but is still discernible in the first three channels ($\sim5$--8 km s$^{-1}$). Therefore, the variation of the line ratio (and thus the excitation temperature) of the I22506 outflow is qualitatively consistent with the model of an episodic jet creating multiple bow-shocks. However, it could not be ruled out with the existing data that the decrease in the line ratio at higher velocities is partly due to the beam dilution effect in our CO (2-1) data, and in that case the CO (2-1) emission is tracing a more compact structure than the CO (1-0) emission at the highest velocities. Future high resolution CO (2-1) observations will provide further insights into this issue. 

In short, all the observations, including our CO (2-1) and IRAC observations, as well as previous CO (1-0) data, all suggest that the molecular outflow in I22506 is driven by jet bow-shocks. The CO (2-1)/(1-0) line ratio and the P-V diagram of the CO (1-0) data show further evidence for the existence of multiple bow-shocks and thus for an episodic nature of the underlying jet.  

\section{Summary}
We present the CSO CO (2-1) and {\it Spitzer} IRAC observations of the molecular outflow in high-mass star-forming region I22506. We also revisit the published CO (1-0) data, and perform a joint analysis of the CO (2-1) and (1-0) observations. 

The bipolar molecular outflow has a mass of $\sim20~M_{\odot}$ for the gas of outflow velocities $\geq5$ km s$^{-1}$. The mass outflow rate may suggest a mass accretion of order $10^{-4}M_{\odot}$ yr$^{-1}$, which is high enough to form a high-mass star. Promising evidence for the jet bow-shocks as the driving agents of the molecular outflow comes from the {\it Spitzer} IRAC image which reveals a pair of bow-shaped tips located at the leading fronts of the bipolar molecular outflow. The derived CO (2-1)/(1-0) line ratio as a function of the outflow velocity, along with the P-V diagram of the high resolution CO (1-0) data, are consistent with the scenario that the outflow is being driven by multiple bow-shocks created by an episodic jet. Thus the mass loss process close to the central protostar in I22506 appears to be a scaled-up version of that is seen in low-mass protostars.

\normalem
\begin{acknowledgements}
This material is based upon work at the Caltech Submillimeter Observatory, which was operated by the California Institute of Technology under cooperative agreement with the National Science Foundation (AST-0838261). This work is based in part on observations made with the Spitzer Space Telescope, which is operated by the Jet Propulsion Laboratory, California Institute of Technology under a contract with the National Aeronautics and Space Administration (NASA). We are grateful to Q. Zhang for providing us the CO (1-0) data. K.Q. acknowledges supports from National Natural Science Foundation of China (NSFC) through grants NSFC 11473011 and NSFC 11590781.

\end{acknowledgements}

\bibliographystyle{raa}
\bibliography{bibtex}

\end{document}